\documentclass[pre, twocolumn, showpacs, preprintnumbers]{revtex4}

\usepackage{amssymb}
\usepackage{graphicx}
\usepackage{amsmath}
\usepackage{bm}
\usepackage{latexsym}
\usepackage{times}
\DeclareMathAlphabet{\mathpzc}{OT1}{pzc}{m}{it}
\DeclareMathAlphabet{\mathcalligra}{T1}{calligra}{m}{n}
\usepackage{setspace}

\usepackage{placeins}

\begin{document}
\preprint{APS/123-QED}

\title{Significance of time-convolutionless mode-coupling theory \\in capturing the dynamics of glass-forming liquids}
\author{Michio Tokuyama}
\address{Professor Emeritus, Institute of Fluid Science, Tohoku University, Sendai 980-8577, Japan}

\date{\today}
\begin{abstract}
This paper demonstrates the significance of the recently proposed time-convolutionless mode-coupling theory (TMCT) in capturing the dynamics of glass-forming liquids. The origin of the primary differences between ideal MCT and TMCT is comprehensively explored from a unified perspective. First, we review two distinct projection operator methods in the Heisenberg picture: the time-convolution (TC) formulation proposed by Mori and the time-convolutionless (TCL) formulation proposed by Tokuyama and Mori. We show that the appropriate choice between these frameworks fundamentally depends on the space-time scales of the relevant variables. In TMCT, the TC formulation is applied to the current density, whereas the TCL formulation is applied to the number density because the latter operates on a significantly longer space-time scale. In contrast, ideal MCT applies the TC formulation to both densities. Consequently, the governing equation in TMCT is time-convolutionless, whereas the ideal MCT equation features a time-convolution form. This fundamental difference significantly affects various physical quantities near the glass transition. Finally, the full TMCT equation is transformed into a simplified recursion equation to facilitate numerical analysis, and the key predictions of TMCT are compared with those of ideal MCT.
\end{abstract}

\maketitle

\begin{figure*}[!t]
\begin{singlespace}
\centering
\includegraphics[width=14.0cm]{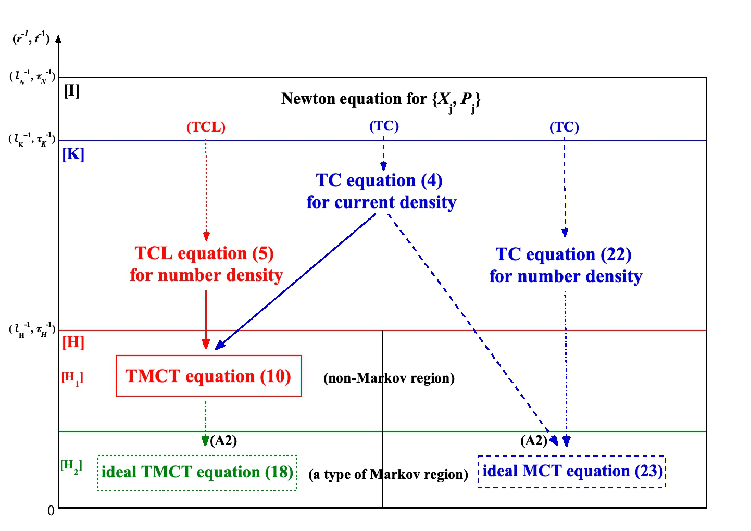}
\caption{(Color online) Classification of the basic equations discussed in the present paper into three stages, [N], [K], and [H], depending on a space ($r$)-time ($t$) scale, where $\ell_i$ and $\tau_i$ indicate the characteristic length and time of interest, respectively. Numbers in the figure indicate equation numbers.}
\label{evo}
\end{singlespace}
\end{figure*}
\section{Introduction}
In 1984, the celebrated mode-coupling theory (MCT) for the glass transition was proposed by Bengtzelius, G\"{o}tze, and Sj\"{o}lander \cite{mct84B}, and independently by Leutheusser \cite{mct84L}. Since then, MCT has provided profound insights into glassy dynamics; however, open questions remain regarding its formalisms and numerical predictions, such as the exact location of the critical point. To determine a physically reasonable critical point, the ideal time-convolutionless mode-coupling theory (ideal TMCT) was introduced in 2014 \cite{toku14}. While the ideal TMCT successfully yields a reasonable critical point, it exhibits inaccurate behavior on intermediate time scales because it is inherently designed for long-term space-time dynamics \cite{toku17}. To resolve this limitation, the full time-convolutionless mode-coupling theory (TMCT) was developed in 2017, which is applicable across all space-time scales except for the microscopic stage [N] governed by Newtonian (or Heisenberg) dynamics. Consequently, TMCT not only reproduces the critical point of the ideal TMCT but also eliminates the unphysical behavior at intermediate times. 

The primary purpose of the present paper is thus to provide a rigorous theoretical proof that explains, from a unified perspective, why TMCT outperforms conventional MCT. Specifically, we elucidate the theoretical foundation of how TMCT accurately captures the dynamics of glass-forming liquids, achieving excellent agreement not only with the transition point but also with numerical simulation results throughout both qualitative and quantitative regimes.

To derive stochastic equations for macrovariables with fluctuating forces from the Newtonian (or Heisenberg) equations, two distinct projection-operator frameworks exist. One is the time-convolution projection-operator method (the so-called TC formulation) proposed by Mori \cite{mori65}, and the other is the time-convolutionless projection-operator method (the so-called TCL formulation) proposed by Tokuyama and Mori \cite{toku75,toku76}. The necessity of these two distinct formulations depends heavily on the space-time scale of the macrovariables. Depending on this scale, the system of interest exhibits three characteristic stages: a microscopic stage [N], a kinetic stage [K], and a hydrodynamic stage [H] (see Fig.~\ref{evo}). The relevant variables required to describe the system dynamics differ in each stage. Notably, the TC formulation is applicable to the relevant variables in stage [K], whereas the TCL formulation is applicable to those in stage [H]. In the microscopic stage [N], the position $\bm{X}_j(t)$ and momentum $\bm{P}_j(t)$ of the $j$-th particle at time $t$ are governed by the Newton (or Heisenberg) equation. 

To illustrate how these two formulations are applied, we present three physical examples. The first example is Brownian motion in an equilibrium liquid. In stage [K], the relevant variables are the velocity $\bm{v}(t) (= d\bm{x}/dt)$ and position $\bm{x}(t) $ of the Brownian particle, whereas only $\bm{x}(t)$ serves as the relevant variable in stage [H]. One can safely apply the TC formulation to $\bm{v}(t)$ to derive the Langevin equation; however, the TCL formulation cannot be applied here because $\bm{v}(t)$ is not a relevant variable in stage [H]. To clarify this situation, we consider the well-known example of the long-time $t^{-3/2}$ tail of the velocity autocorrelation function, which has been experimentally observed \cite{pusey81,hu11}. As demonstrated in Ref.~\cite{toku78}, such a tail is readily obtained by employing the TC formulation, whereas the TCL formulation fails to capture it. Thus, the TCL formulation is not applicable to velocity dynamics. 

The second example is the derivation of the Boltzmann equation from the Heisenberg equation. In stage [K], the relevant variable is the coarse-grained particle density $n(\bm{x},\bm{p};t)$ in $\mu$-space, where $\bm{x}$ and $\bm{p}$ denote the position and momentum of the particle at time $t$, respectively. This variable persists even into stage [H] as a hydrodynamic variable; hence, the TCL formulation is appropriate. As shown in Ref.~\cite{toku761}, employing the TCL formulation successfully derives the Boltzmann equation with the correct Boltzmann collision integral \cite{cohen}. Conversely, the TC formulation fails to properly describe this integral \cite{mori73}.

The final example is the dynamics of glass-forming liquids. In stage [K], the relevant variables are given by the number densities \(\rho_{\alpha}(\bm{q},t)\) and the current densities $\bm{j}_{\alpha}(\bm{q},t) (\propto d\rho_{\alpha}(\bm{q},t)/dt)$, where $\alpha=c$ and $\alpha=s$ denote the collective and self cases, respectively, and $\bm{q}$ represents the wave vector. Here, we note that the space-time regime over which the number densities play a dominant role is broader than that of the current densities. Consequently, in stage [H], the relevant variables are reduced to $\rho_{\alpha}(\bm{q},t)$. The TC formulation is applied to the dynamics of the current densities because their dynamics governs the system exclusively in stage [K]. This yields the linear non-Markovian stochastic equations for $\bm{j}_{\alpha}(\bm{q},t)$ derived from the Newton equation (see the center dashed arrow (TC) in Fig.~\ref{evo}), where the memory terms are written as time convolutions involving the correlation functions of the fluctuating forces. Conversely, the TCL formulation is employed for the dynamics of the number densities, as their dynamics begins to dominate the system in stage [H]. This leads to the linear non-Markovian stochastic equations for $\rho_{\alpha}(\bm{q},t)$ derived from the Newton equation (see the dotted arrow (TCL) in Fig.~\ref{evo}), where the memory terms are time-convolutionless and expressed via the correlation functions of the fluctuating currents.

In stage [H], the system is described by the diffusion equation for the intermediate scattering function $f_{\alpha}(q,t)$, which is defined by the density-density correlation function. To obtain closed equations for $f_{\alpha}(q,t)$, the coupled TC equations for $\bm{j}_{\alpha}(\bm{q},t)$ and TCL equations for $\rho_{\alpha}(\bm{q},t)$ are utilized. This framework successfully derives the time-convolutionless mode-coupling theory (TMCT) equations for $f_{\alpha}(q,t)$ previously proposed in Ref.~\cite{toku17} (see the arrow and down-left arrow on the left in Fig.~\ref{evo}). Depending on the space-time scale, the hydrodynamic stage [H] further comprises two substages: a fast (or non-Markovian) stage [H$_1$] and a slow (or Markovian-type) stage [H$_2$]. The TMCT equations for $f_{\alpha}(q,t)$ hold in stage [H$_1$]. In stage [H$_2$], the ideal TMCT equations developed in Refs.~\cite{toku14,toku15} are readily recovered from the full TMCT equations by applying the Approximation (A2) (a type of Markov approximation) discussed in Ref.~\cite{toku14} (see the final dot-dashed arrow with (A2) on the left in Fig.~\ref{evo}).

These examples confirm that to derive stochastic equations from the microscopic framework, the TC formulation is applicable to macrovariables that serve as relevant variables only in stage [K], whereas the TCL formulation must be applied to macrovariables that remain relevant across both stages [K] and [H].

We next examine the conventional ideal mode-coupling theory (ideal MCT) \cite{mct84B,mct84L} from the novel perspective established above. Crucially, in the ideal MCT, only the TC formulation was applied to derive the basic equations for both relevant variables in stage [K] (see the center and right-hand dashed arrows (TC) in Fig.~\ref{evo}). Consequently, while the resulting linear non-Markovian stochastic equations for $\bm{j}_{\alpha}(\bm{q},t)$ are identical to those obtained in TMCT, the corresponding equations for $\rho_{\alpha}(\bm{q},t)$ differ from the TMCT framework because their memory terms retain a time-convolution form. In stage [H$_2$], these coupled equations are utilized in conjunction with the Approximation (A2) to yield the conventional ideal MCT equations for $f_{\alpha}(q,t)$ (see the down-right dashed arrow with (A2) and the dot-dashed arrow on the right in Fig.~\ref{evo}).

The ideal MCT equations have been numerically solved for various glass-forming systems \cite{mct91,fuch,fuchs,fran,chong,voigt03,got03,sza,foffi,voigt04,flenn,voigt06,toku08,got09,voigt10,narumi11,voigt11}. Although the full numerical solutions of these equations successfully predict an ergodic-to-non-ergodic transition at a critical temperature $T_c$ (or a critical volume fraction $\phi_c$), the predicted $T_c$ (or $\phi_c$) is systematically higher (or lower) than the thermodynamic glass transition temperature $T_g$ (or $\phi_g$), which is conventionally defined by the crossover point in an enthalpy-temperature curve \cite{debe}. To overcome this "high-$T_c$" problem, the ideal TMCT equations for $f_{\alpha}(q,t)$ were proposed in Refs.~\cite{toku14,toku15} by employing a formulation structurally analogous to MCT, with the pivotal exception that the TCL formulation is applied to the densities instead of the TC formulation.

Subsequent studies \cite{toku15,toku171} within the simplified model introduced by MCT demonstrated that the ideal TMCT equations yield non-zero long-time solutions for $T \leq T_c$, where the revised $T_c$ is significantly lower than that predicted by MCT. Furthermore, as a preliminary test of the ideal TMCT, the equations were solved numerically \cite{kimu14,kimura16} using the Percus-Yevick static structure factor for hard spheres \cite{PY} under the exact conditions employed by Chong et al. \cite{chong} for the ideal MCT. Concurrently, the MCT equations were re-solved, and their numerical results agreed with those obtained by Voigtmann et al. \cite{voigt04} within error margins. These comparative analyses revealed that the critical volume fraction $\phi_c$ of the ideal TMCT is substantially higher than that of the ideal MCT, aligning closely with the glass transition volume fraction $\phi_g$ predicted by hard-sphere simulations \cite{toku03,toku031}. This agreement strongly demonstrates that the TC formulation is inapplicable to the dynamics of number densities in stage [H$_2$].

Additionally, the ideal TMCT equations were extended in Ref.~\cite{toku191} to derive ion transport equations under a weak electric field. The resulting analytical ionic conductivity exhibits a striking departure from the well-known Nernst-Einstein relation. Currently, only the qualitative behavior of these analytical results has been validated by experiments on the ionic liquid PC-LiPF$_6$, because the static structure factor for this specific liquid system remains unavailable both analytically and numerically.

Crucially, employing the TCL formulation for the number densities is indispensable to recover the cumulant expansion formulated by Kubo \cite{kubo62}. TMCT thus enables the consistent calculation of each cumulant expansion term from first principles, including the mean-square displacement $M_2(t)$ and the non-Gaussian parameter $\alpha_2(t) \equiv 3M_4(t)/5M_2(t)^2 - 1$ \cite{ng}, where $M_{2n}(t)=\langle|\bm{X}_j(t)-\bm{X}_j(0)|^{2n}\rangle$ and the brackets denote the average over an equilibrium ensemble. Physically, $\alpha_2(t)$ must satisfy $\alpha_2(t=0) = \alpha_2(t=\infty) = 0$ and remain non-negative ($\alpha_2(t) \geq 0$) for all times. As demonstrated in Ref.~\cite{toku17}, however, the Approximation (A2) induces unphysical behavior in these physical quantities at intermediate time scales, specifically during the $\beta$-relaxation stage. Indeed, the ideal TMCT unphysically yields $\alpha_2(t) < 0$ in the $\beta$ stage. The conventional ideal MCT also yields $\alpha_2(t) < 0$ during the $\beta$ stage, accompanied by the unphysical initial baseline value $\alpha_2(t=0) = -2/3$. The origin of this artifact ($\alpha_2(t) < 0$) stems entirely from the application of the Approximation (A2), whereas the unphysical initial condition $\alpha_2(t=0) = -2/3$ is inherently tied to the time-convolution equation governing $f_{\alpha}(q,t)$.

As shown in Ref.~\cite{toku17}, the full TMCT equations successfully eliminate these unphysical artifacts. To render these fundamental equations more amenable to numerical computation, the present paper transforms them into simpler recursion equations without invoking the Approximation (A2), strictly adhering to the exact formulation employed in the original derivation of the TMCT equations.

\section{Basic equations}
In the present section, we briefly summarize the basic equations obtained by TMCT and also those by MCT for comparison. We consider the three-dimensional equilibrium glass-forming system, which consists of $N$ particles with mass $m$ and diameter $\sigma$ in the total volume $V$ at temperature $T$.  Let $\xi$ denote the control parameter, such as a volume fraction $\phi(=\pi\rho \sigma^3/6)$ and an inverse temperature $1/T$, where $\rho(=N/V)$ is the number density. 

\subsection{Coupled equations in stage [K]}
We first discuss the equations in stage [K]. The relevant variables are the number density fluctuations $\rho_{\alpha}(\bm{q},t)$ and the current density fluctuations $j_{\alpha}(\bm{q},t)$ given by
\begin{equation}
\begin{split}
\rho_c(\bm{q},t)&=\frac{1}{N^{1/2}}\left[\sum_{j=1}^N\rho_s(\bm{q},t)-N\delta_{\bm{q},0}\right],\\
 \rho_s(\bm{q},t)&=e^{i\bm{q}\cdot\bm{X}_j(t)},
\end{split}
 \label{rhos}
\end{equation}
\begin{equation}
\begin{split}
j_c(\bm{q},t)&=\frac{1}{N^{1/2}}\sum_{j=1}^N \hat{\bm{q}}\cdot\frac{\bm{p}_j(t)}{m} e^{i\bm{q}\cdot\bm{X}_j(t)},\\
j_s(\bm{q},t)&= \hat{\bm{q}}\cdot\frac{\bm{p}_j(t)}{m} e^{i\bm{q}\cdot\bm{X}_j(t)},
\end{split}
\label{currs}
\end{equation}
respectively, where $\langle\rho_s(\bm{q},t)\rangle=\delta_{\bm{q},0}$, and $\langle\rho_c(\bm{q},t)\rangle=\langle j_{\alpha}(\bm{q},t)\rangle=0$. Since these fluctuations are macroscopic physical quantities, we set $0 \leq q\leq q_c$, where the inverse cutoff $q_c^{-1}$ is longer than a linear range of the intermolecular force but shorter than a semi-macroscopic length and is in general fixed so that the numerical solutions coincide with the simulation results or the experimental data at least in a liquid state \cite{toku17}. 
The scaled intermediate scattering function $f_{\alpha}(q,t)$ is then given by
\begin{equation}
f_{\alpha}(q,t)=\langle \rho_{\alpha}(\bm{q},t)\rho_{\alpha}(\bm{q},0)^*\rangle/S_{\alpha}(q) \label{sfa}
\end{equation}
with $S_c(q)=S(q)$ and $S_s(q)=1$, where the static structure factor $S(q)$ is given by $S(q)=\langle |\rho_c(\bm{q},0)|^2\rangle$, $q=|\bm{q}|$, and the asterisk denotes the complex conjugate. As shown in the previous paper \cite{toku17}, by applying the TCL type for $\rho_{\alpha}(\bm{q},t)$, and the TC type for $j_{\alpha}(\bm{q},t)$, one can find the coupled equations for the current-current correlation function $\psi_{\alpha}(\bm{q},t)$ and for the scattering function $f_{\alpha}(\bm{q},t)$ as
\begin{equation}
\begin{split}
\frac{\partial}{\partial t}\psi_{\alpha}(q,t)&=-\gamma_{\alpha}\psi_{\alpha}(q,t)\\
&- \int_0^t\Delta\varphi_{\alpha}(q,t-s)\frac{f_{\alpha}(q,s)}{f_{\alpha}(q,t)}\psi_{\alpha}(q,s)ds,
\end{split}
\label{eqpsi0}
\end{equation}
\begin{equation}
\begin{split}
\frac{\partial}{\partial t}f_{\alpha}(q,t)&=-q^2\left[\int_0^t \psi_{\alpha}(q,s)ds \right]f_{\alpha}(q,t)\\
&\equiv -q^2D_{\alpha}(q,t)f_{\alpha}(q,t)
\end{split}
\label{eqf}
\end{equation}
with the nonlinear memory function
\begin{equation}
\begin{split}
\Delta\varphi_{\alpha}(q,t)&=\frac{v_{th}^2}{2^{n_{\alpha}}\rho}\int_<\frac{d\bm{k}}{(2\pi)^3}v_{\alpha}(\bm{q},\bm{k})^2S(k)S_{\alpha}(|\bm{q}-\bm{k}|)\\
&\times f_c(k,t)f_{\alpha}(|\bm{q}-\bm{k}|,t),
\end{split}
\label{memory}
\end{equation}
where $\gamma_{\alpha}$ is a damping constant to be determined by the fitting with the simulation results or the experimental data and depends only on the intensive parameter, such as the volume fraction $\phi$ \cite{toku17}, and $v_{th}(=(k_BT/m)^{1/2})$ an average thermal velocity. Here $D_{\alpha}(q,t)$ is the time- and q-dependent diffusion coefficient, and $\int_<$ denotes the sum over wave vectors $\bm{k}$ whose magnitudes are smaller than a cutoff $q_c$. The vertex amplitude $v_{\alpha}(\bm{q},\bm{k})$ is given by 
\begin{equation}
v_{\alpha}(\bm{q},\bm{k})=\hat{\bm{q}}\cdot\bm{k}c(k)+n_{\alpha}\hat{\bm{q}}\cdot(\bm{q}-\bm{k})c(|\bm{q}-\bm{k}|),\label{vertex}
\end{equation}
where $\rho c(k)=1-1/S(k)$, $\hat{\bm{q}}=\bm{q}/q$, and $n_c=1$ and $n_s=0$. The initial conditions are given by $f_{\alpha}(q,0)=1$ and $\psi_{\alpha}(q,0)=v_{th}^2/S_{\alpha}(q)$. It is worth noting that the equation (\ref{eqf}) for $f_{\alpha}(q,t)$ is time-convolutionless in time, whereas the equation (\ref{eqpsi0}) for $\psi_{\alpha}(q,t)$ involves a time convolution. Further more, the nonlinear memory function $\Delta\varphi_{\alpha}(q,t)$ has exactly the same form as that obtained in MCT \cite{mct84B,mct84L,mct91}. This is reasonable because the TC type is applied to $j_{\alpha}(\bm{q},t)$ in both theories. As will be discussed later, the difference between TMCT and MCT appears in the equations for $f_{\alpha}(q,t)$ since different formulations, namely the TCL type and TC type, are applied to $\rho_{\alpha}(\bm{q},t)$. Finally, we also emphasize that as shown in Ref. \cite{toku18}, the same equations as Eqs. (\ref{eqpsi0}) and  (\ref{eqf}) hold even for colloidal suspensions, except that the nonlinear memory function $\Delta\varphi_{\alpha}(q,t)$ contains the hydrodynamic correlation effect in addition to the mechanical correlation effect. 

\subsection{TMCT equations in stage [H]}
In stage [H], the relevant variables are given by the density fluctuations $\rho_{\alpha}(\bm{q},t)$. To obtain a single closed equation for $f_{\alpha}(q,t)$ from the coupled equations (\ref{eqpsi0}) and (\ref{eqf}), we first introduce the cumulant function $K_{\alpha}(q,t)$ via
\begin{equation}
f_{\alpha}(q,t)=\exp[-K_{\alpha}(q,t)], \label{fsol}
\end{equation}
which is combined with Eq.~(\ref{eqf}) to yield
\begin{equation}
K_{\alpha}(q,t)=q^2 \int_0^t D_{\alpha}(q,s) ds =q^2\int_0^t (t-s)\psi_{\alpha}(q,s)ds. \label{cumu}
\end{equation}
From Eq.~(\ref{cumu}), one can readily find the relations $\partial K_{\alpha}(q, t)/\partial t=q^2D_{\alpha}(q, t)$ and $\partial^2K_{\alpha}(q, t)/\partial t^2=q^2\psi_{\alpha}(q, t)$. 

As shown in Ref.~\cite{toku17}, integrating Eq.~(\ref{eqpsi0}) over $t$ and using the initial condition $\psi_{\alpha}(q, 0) = v_{th}^2/S_{\alpha}(q)$, we obtain the following second-order differential equation for $K_{\alpha}(q, t)$: 
\begin{equation}
\begin{split}
&\frac{\partial^2 K_{\alpha}(q,t)}{\partial t^2} = \frac{q^2v_{th}^2}{S_{\alpha}(q)}-\gamma_{\alpha}\frac{\partial K_{\alpha}(q,t)}{\partial t} \\
&- \int_0^tds \int_0^s d\tau\Delta\varphi_{\alpha}(q,s-\tau)\frac{f_{\alpha}(q,\tau)}{f_{\alpha}(q,s)}\frac{\partial^2 K_{\alpha}(q,\tau)}{\partial \tau^2}, 
\end{split}
\label{Kcumu}
\end{equation}
where the initial conditions are given by $K_{\alpha}(q,0)=0$ and $\partial K_{\alpha}(q,t)/\partial t |_{t=0}=0$. Equation~(\ref{Kcumu}) is the TMCT equation for $K_{\alpha}(q,t)$ derived in Ref.~\cite{toku17} to describe the dynamics of glass-forming materials in stage [H]. 

It is worth noting that the approximation (A2) is not employed in deriving Eq.~(\ref{Kcumu}). This equation has been solved numerically in Ref.~\cite{toku17}, where it was demonstrated that the numerical solutions do not exhibit any unphysical behavior for intermediate times. In what follows, we show how to transform Eq.~(\ref{Kcumu}) into a simpler equation by employing the Approximation (A1), defined as $|\psi_{\alpha}(q,t)D_{\alpha}(q,t)|\ll 1$, which was also used to derive Eq.~(\ref{eqpsi0}).

\subsection{TMCT recursion equations in stage [H]}
By using the approximation (A1) discussed in Ref.~\cite{toku17}, we next derive a simpler equation from Eq.~(\ref{Kcumu}). We first integrate Eq.~(\ref{Kcumu}) over $t$ and then evaluate the memory term by performing integration by parts and employing the approximation (A1). As shown in Appendix A, after detailed calculations, one finally obtains
\begin{equation}
\begin{split}
&K_{\alpha}(q,t) = \frac{q^2v_{th}^2}{S_{\alpha}(q)}C_{\alpha}(t) \\
&- \int_0^{t}ds e^{-\gamma_{\alpha}(t-s)}\int_0^{s}d\tau\Delta\varphi_{\alpha}(q,s-\tau)\frac{f_{\alpha}(q,\tau)}{f_{\alpha}(q,s)}K_{\alpha}(q,\tau),
\end{split}
\label{K0f}
\end{equation}
where
\begin{equation}
C_{\alpha}(t)=(\gamma_{\alpha} t-1+e^{-\gamma_{\alpha} t})/\gamma_{\alpha}^2.\label{Ca}
\end{equation}
Equation~(\ref{K0f}) is the TMCT recursion equation used to discuss the dynamics of glass-forming materials in stage [H]. Here, we should emphasize that it does not cause any unphysical behavior for intermediate times because the approximation (A2) is not employed in its derivation. In fact, the non-Markovian term given by $f_{\alpha}(q,\tau)/f_{\alpha}(q,s)$ in Eq.~(\ref{K0f}) ensures the accurate behavior of physical quantities for intermediate times \cite{toku17}.

To proceed, it is convenient to introduce the Laplace transform of $K_{\alpha}(q, t)$, defined as $K_{\alpha}[q,z]=\mathcal{L}[K_{\alpha}(q,t)][z]:= \int_0^{\infty}e^{-zt}K_{\alpha}(q,t)dt$. Then, from Eq.~(\ref{K0f}), one finds
\begin{equation}
K_{\alpha}[q,z]=\frac{q^2v_{th}^2}{S_{\alpha}(q)}\frac{1}{z^2(z+\gamma_{\alpha}+M_{\alpha}[q,z])}, \label{Kzr}
\end{equation}
where the memory term $M_{\alpha}[q,z]$ is given by
\begin{equation}
\begin{split}
M_{\alpha}[q,z] &= \frac{1}{K_{\alpha}[q,z]}\int_0^{\infty}ds e^{-zs}\Delta\varphi_{\alpha}(q,s) \\
&\times \int_0^{\infty}dt e^{-zt}\frac{f_{\alpha}(q,t)}{f_{\alpha}(q,s+t)}K_{\alpha}(q,t).
\end{split}
\label{mz}
\end{equation}

From Eq.~(\ref{K0f}) [or Eq.~(\ref{mz})], one can determine the asymptotic forms of $K_{\alpha}(q,t)$. For short times $t \ll 1/\gamma_{\alpha}$ (or $z \gg \gamma_{\alpha}$), we obtain the following expression for an arbitrary value of $\xi$:
\begin{equation}
K_{\alpha}(q,t)\simeq \frac{q^2v_{th}^2}{2S_{\alpha}(q)}t^2, \quad K_{\alpha}[q,z]\simeq \frac{q^2v_{th}^2}{S_{\alpha}(q)z^3}. \label{K0}
\end{equation}
In the long-time limit $t \rightarrow \infty$ (or $z \rightarrow 0$), we obtain for $\xi < \xi_c$:
\begin{equation}
K_{\alpha}(q,t) \simeq q^2D_{\alpha}(q)t, \quad K_{\alpha}[q,z] \simeq \frac{q^2D_{\alpha}(q)}{z^2}, \label{Kin}
\end{equation}
where the $q$-dependent diffusion coefficient is given by
\begin{equation}
D_{\alpha}(q) = D_{\alpha}(q,t=\infty) = \frac{v_{th}^2/S_{\alpha}(q)}{\gamma_{\alpha}+\int_0^{\infty}\Delta\varphi_{\alpha}(q,s)ds}, \label{Dal}
\end{equation}
and $\xi_c$ is the critical value of $\xi$ discussed later.

\subsection{Ideal TMCT equations in stage [H$_2$]}
In the slow stage [H$_2$], by applying the approximation (A2) discussed in Ref.~\cite{toku17} to Eq.~(\ref{Kcumu}), one can further obtain the following equation for $K_{\alpha}(q,t)$:
\begin{equation}
\begin{split}
\frac{\partial^2 K_{\alpha}(q,t)}{\partial t^2} &= \frac{q^2v_{th}^2}{S(q)}-\gamma_{\alpha}\frac{\partial K_{\alpha}(q,t)}{\partial t}\\
 &- \int_0^t\Delta\varphi_{\alpha}(q,t-s)\frac{\partial K_{\alpha}(q,s)}{\partial s}ds. 
 \end{split}
\label{Kcumu0}
\end{equation}
Equation~(\ref{Kcumu0}) is the ideal TMCT equation already discussed in Refs.~\cite{toku14,toku15,toku171}. By solving this equation formally, one can find
\begin{equation}
\begin{split}
K_{\alpha}(q,t) &= \frac{q^2v_{th}^2}{S_{\alpha}(q)}C_{\alpha}(t) \\
&- \int_0^{t}ds e^{-\gamma_{\alpha}(t-s)}\int_0^{s}d\tau\Delta\varphi_{\alpha}(q,s-\tau)K_{\alpha}(q,\tau). 
\end{split}
\label{K0f0}
\end{equation}
Equation~(\ref{K0f0}) represents the ideal TMCT recursion equation for $K_{\alpha}(q,t)$. Here, we note that Eq.~(\ref{K0f0}) can also be straightforwardly derived by directly applying the approximation (A2) to Eq.~(\ref{K0f}). Based on this equation, two types of relaxation dynamics in the $\alpha$ and $\beta$ stages near the critical point have been successfully analyzed \cite{kimura16} by employing a similar approach to that used in MCT \cite{toku192}. However, we should mention here that as discussed in Refs.~\cite{toku17,toku15}, the numerical solutions of Eq.~(\ref{Kcumu0}) do not coincide with those of Eq.~(\ref{Kcumu}) in the $\beta$ stage due to the limitations of the approximation (A2), although they agree with each other at short and long times. 

By taking the Laplace transform of Eq.~(\ref{K0f0}), one obtains
\begin{equation}
K_{\alpha}[q,z]=\frac{q^2v_{th}^2}{S_{\alpha}(q)}\frac{1}{z^2(z+\gamma_{\alpha}+\Delta\varphi_{\alpha}[q,z])}. \label{Kzi}
\end{equation}
We remark that the asymptotic forms of $K_{\alpha}(q,t)$ derived from this ideal version are identical to those of the full TMCT.

\subsection{Ideal MCT equation in stage [H$_2$]}
Finally, we discuss the MCT equations for $f_{\alpha}(q,t)$ from a unified point of view mentioned in the Introduction. They are derived from the coupled TC equations for both the current and number densities (see the center and right-hand dashed arrows (TC) in Fig.~\ref{evo}). In fact, by applying the TC formulation to the current densities, one can obtain the following TC equations for $\psi_{\alpha}(q,t)$ \cite{toku14}:
\begin{equation}
\frac{\partial}{\partial t}\psi_{\alpha}(q,t)=-\gamma_{\alpha}\psi_{\alpha}(q,t)
- \int_0^t\Delta\varphi_{\alpha}(q,t-s)\psi_{\alpha}(q,s)ds, \label{eqpsi0m}
\end{equation}
where the approximation (A2) is also employed in the derivation. Here, we note that Eq.~(\ref{eqpsi0m}) can be directly derived from Eq.~(\ref{eqpsi0}) by using the approximation (A2), where $f_{\alpha}(q,s)\simeq f_{\alpha}(q,t)$. 

By applying the TC formulation to the number densities, one can also derive the following TC equations for $f_{\alpha}(q,t)$ \cite{toku14}:
\begin{equation}
\frac{\partial f_{\alpha}(q,t)}{\partial t}=- \int_0^t \psi_{\alpha}(q,t-s)f_{\alpha}(q,s)ds.\label{mct0}
\end{equation}
It should be emphasized that Eq.~(\ref{mct0}) is quite different from Eq.~(\ref{eqf}), although the memory functions in both theories coincide with each other within the MCT binary approximation \cite{mct91}. Combining the coupled equations (\ref{eqpsi0m}) and (\ref{mct0}) leads to
\begin{equation}
\begin{split}
\frac{\partial^2f_{\alpha}(q,t)}{\partial t^2} &= -\frac{q^2v_{th}^2}{S_{\alpha}(q)}f_{\alpha}(q,t)-\gamma_{\alpha}\frac{\partial f_{\alpha}(q,t)}{\partial t} \\
&- \int_0^t\Delta\varphi_{\alpha}(q,t-s)\frac{\partial f_{\alpha}(q,s)}{\partial s}ds. 
\end{split}
\label{mct}
\end{equation}
This is the ideal MCT equation discussed in Ref.~\cite{mct84B}. By taking the Laplace transform of Eq.~(\ref{mct}), one finds
\begin{equation}
f_{\alpha}[q,z]=\left(z+\frac{q^2v_{th}^2}{S_{\alpha}(q)}\frac{1}{z+\gamma_{\alpha}+\Delta\varphi_{\alpha}[q,z]}\right)^{-1}. \label{Kzmct}
\end{equation}
We remark that the asymptotic forms of $f_{\alpha}(q,t)$ derived from MCT are identical to those of TMCT. 

As shown in Refs.~\cite{kimura16,kimu14}, Eqs.~(\ref{Kcumu0}) and (\ref{mct}) have already been solved numerically based on the Percus-Yevick static structure factor for hard spheres \cite{PY} under the same conditions as those employed by Chong et al. \cite{chong} to solve the ideal MCT equations. Thus, the characteristic differences between ideal TMCT and ideal MCT have also been well elucidated.

\section{An ergodic to non-ergodic transition}
The most important prediction of MCT is the ergodic to non-ergodic transition at a critical point $\xi_c$, above which the solution $f_{\alpha}(q,t)$ reduces to a non-zero value for long times. In fact, from Eq.~(\ref{mct}), one can find for $\xi\geq \xi_c$ \cite{mct84B}
\begin{equation}
f_{\alpha}(q)=\lim_{t\rightarrow \infty}f_{\alpha}(q,t)=\frac{\mathpzc{F}_{\alpha}(q)}{1+\mathpzc{F}_{\alpha}(q)}
\label{mctfq}
\end{equation}
with the long-time limit of the nonlinear memory function
\begin{equation}
\begin{split}
&\mathpzc{F}_{\alpha}(q,f_c,f_{\alpha}) = \lim_{z\rightarrow 0}\frac{z\Delta\varphi_{\alpha}[q,z]}{q^2v_{th}^2}S_{\alpha}(q) \\
&= \frac{1}{2^{n_{\alpha}}(2\pi)^3}\int_< d\bm{k}V_{\alpha}^{(2)}(q, k, |\bm{q}-\bm{k}|) f_c(k)f_{\alpha}(|\bm{q}-\bm{k}|),
\end{split}
\label{Ga1}
\end{equation}
and the vertex
\begin{equation}
V_{\alpha}^{(2)}(q, k, |\bm{q}-\bm{k}|)=S_{\alpha}(q)S_c(k)S_{\alpha}(|\bm{q}-\bm{k}|)v_{\alpha}(\bm{q},\bm{k})^2/(\rho q^2).\label{V}
\end{equation}
Here, the non-zero value $f_{\alpha}(q)$ is the so-called non-ergodicity parameter.  

As shown in previous papers \cite{toku14,toku17}, this prediction also holds for TMCT. In fact, from Eq.~(\ref{K0f}) [or Eq.~(\ref{K0f0})], one can find for $\xi\geq \xi_c$ 
\begin{equation}
K_{\alpha}(q)=\lim_{t\rightarrow \infty}K_{\alpha}(q,t)=\frac{1}{\mathpzc{F}_{\alpha}(q)}, \label{ene}
\end{equation}
which leads to 
\begin{equation}
f_{\alpha}(q)=\lim_{t\rightarrow \infty}f_{\alpha}(q,t)=\exp\left[- \frac{1}{\mathpzc{F}_{\alpha}(q)}\right].\label{fq}
\end{equation}
The existence of the critical point in Eq.~(\ref{ene}) is mathematically confirmed since $K_c(q)$ is closely related to the Lambert W-function. 

To estimate how the TMCT critical point obtained by Eq.~(\ref{fq}) differs from that of MCT obtained by Eq.~(\ref{mctfq}), it is convenient to employ the simplified model discussed by Bengtzelius et al. \cite{mct84B}. Thus, we can approximate $S_c(q)$ as $S_c(q)=1+A\delta(q-q_m)$, where $A$ is a positive constant to be determined and $q_m$ is the wave vector of the first peak of $S_c(q)$. Then, Eq.~(\ref{Ga1}) can be written as $\mathpzc{F}_c(q_m)=\lambda' f_c(q_m)^2$, where the coupling parameter $\lambda'$ is given by $\lambda'=q_mA^2S(q_m)/(8\pi^2\rho)$. 

Using Eqs.~(\ref{mctfq}) and (\ref{Ga1}) leads to the critical coupling parameter $\lambda'_c=4$ and $f_c=1/2$ for MCT, whereas using Eqs.~(\ref{Ga1}) and (\ref{fq}) yields $\lambda'_c=2e \ (\simeq 5.43656)$ and $f_c=e^{-1/2}$ for TMCT. Thus, the critical coupling parameter of TMCT is larger than that of MCT, which suggests that the critical temperature $T_c$ (or the critical volume fraction $\phi_c$) of TMCT is much lower (or higher) than that of MCT. Indeed, this behavior has been confirmed in Refs.~\cite{kimura16,kimu14} by solving the ideal TMCT and MCT equations numerically based on the Percus-Yevick static structure factor, where $\gamma_{\alpha}$ is set to zero for simplicity. Consequently, the critical volume fraction $\phi_c$ of TMCT is found to be much higher than that of MCT. 

Remarkably, the critical volume fraction of TMCT agrees within error with simulation results for monodisperse hard spheres \cite{toku03,toku031,toku05,toku071}. A similar situation was reported in previous papers \cite{toku141,toku19}, where the critical point $\xi_c$ ($T_c$, $\phi_c$) was shown to be close to the so-called thermodynamic glass transition point $\xi_g$ ($T_g$, $\phi_g$) from the asymptotic solutions of the ideal TMCT equations within the simplified model. 

Equations (\ref{mctfq}) and (\ref{fq}) hold even for colloidal suspensions. In fact, they have been used to study the phase diagram of the kinetic glass transition for short-range attractive colloids \cite{daw,naru17}. Under these conditions, TMCT has been shown to recover all the remarkable features predicted by MCT for attractive colloids while significantly improving the predicted critical values of $\phi_c$.

\section{Asymptotic behavior of $f_c(q,t)$}
We here discuss the asymptotic behavior of $f_c(q,t)$.

\subsection{A two-step relaxation process in a $\beta$ stage}
The second important prediction of MCT is the existence of a two-step relaxation process in the $\beta$ stage. As demonstrated in Refs.~\cite{mct84B,mct91}, MCT shows that $f_c(q,t)$ obeys a characteristic two-step relaxation process at the so-called $\beta$-relaxation stage [$\beta$] near the critical point. By taking the Laplace transform of Eq.~(\ref{mct}), one obtains
\begin{equation}
\frac{zf_c[q,z]}{1-zf_c[q,z]}=z\mathcal{L}[\mathpzc{F}_c(q,f_c(t),f_c(t))][z].\label{mctz}
\end{equation}
Following the standard MCT framework \cite{mct91}, one can split $f_c(q,t)$ into a trivial asymptotic part and a non-trivial part $G(t)$:
\begin{equation}
f_c(q,t)=f_c^c(q)+h_qG(t),\quad zf_c[q,z]=f_c^c(q)+zh_qG[z], \label{split}
\end{equation}
with $h_q=(1-f_c^c(q))^2e_q^c$, where $e_q^c$ is an appropriately normalized right eigenvector of the stability matrix $C_{qk}=(\partial \mathpzc{F}_c/ \partial f_c(k))(1-f_c^c(k))^2$ at $\xi_c$. From Eqs.~(\ref{mctz}) and (\ref{split}), one can find near $\xi_c$:
\begin{equation}
\sigma'+\lambda_{mct} \{z\mathcal{L}[G(t)^2][z]\}-\{zG[z]\}^2=0, \label{sol}
\end{equation}
where $\sigma'$ is a separation parameter given by $\sigma'=c'(\xi/\xi_c-1)$, $c'$ being a positive constant to be determined. Here, $\lambda_{mct}$ is the exponent parameter given by
\begin{equation}
\lambda_{mct}=\frac{1}{2}\sum_{q,k,p}\hat{e}_q^cV_c^{(2)}(q,k,p)h_kh_p, \label{lamdam}
\end{equation}
where $\hat{e}_q^c$ is a left eigenvector defined by $\sum_q\hat{e}_q^ce_q^c=1$, and $p=|\bm{q}-\bm{k}|$. Solving Eq.~(\ref{sol}) leads to two different power-law decays for $G(t)$ near $\xi_c$: the critical decay at the fast $\beta$ stage,
\begin{equation}
G(t)=|\sigma'|^{1/2}(t_{\sigma}/t)^a, \quad t_0\ll t\leq t_{\sigma}, \label{critical}
\end{equation}
and the von Schweidler decay at the slow $\beta$ stage,
\begin{equation}
G(t)=-(t/t_{\sigma}')^b, \quad t_{\sigma}\leq t \ll t_{\sigma}', \label{von}
\end{equation}
where $t_0$ is a microscopic time, $t_{\sigma}=t_0|\sigma'|^{-1/2a}$, and $t_{\sigma}'=t_0B^{-1/b}|\sigma'|^{\gamma}$. Here, $\gamma=1/2a+1/2b$ and $B$ is a positive constant to be determined. The time exponents $a$ and $b$ are determined through the relation
\begin{equation}
\frac{\Gamma(1-a)^2}{\Gamma(1-2a)}=\frac{\Gamma(1+b)^2}{\Gamma(1+2b)}=\lambda_{mct}, \label{ab}
\end{equation}
where $\Gamma(x)$ denotes the Gamma function. For the Percus-Yevick (PY) model, $\lambda_{mct}$ is calculated as $\lambda_{mct}=0.7349$ at $q_c\sigma=40$, which yields $a=0.3117$ and $b=0.5830$ \cite{fran}.

The above mathematical formulation can be directly applied to the TMCT equation \cite{got15,toku192}. We first discuss the ideal TMCT. Utilizing Eq.~(\ref{Kzi}) leads to
\begin{equation}
\frac{1}{zK_c[q,z]}=z\mathcal{L}[\mathpzc{F}_c(q,f_c(t),f_c(t))][z]. \label{itmctz}
\end{equation}
We first write
\begin{equation}
\begin{split}
K_c(q,t) &= K_c^c(q)[1-G(t)], \\
zK_c[q,z] &= K_c^c(q)[1-zG[z]],
\end{split}
\label{split3}
\end{equation}
where $K_c^c=-\ln(f_c^c)$. Use of Eq.~(\ref{fsol}) then leads to, to the lowest order in $K_c^c$:
\begin{equation}
\begin{split}
f_c(q,t) &= f_c^c(q)+K_c^c(q)G(t), \\
zf_c[q,z] &= f_c^c(q)+zK_c^c(q)G[z].
\end{split}
\label{split2}
\end{equation}
From Eq.~(\ref{itmctz}), therefore, the same equation as Eq.~(\ref{sol}) is found, except that $\lambda_{mct}$ is replaced by $\lambda_i$ which is given by
\begin{equation}
\lambda_i=\frac{1}{2}\sum_{q,k,p}K_c^c(q)V_c^{(2)}(q,k,p)K_c^c(k)K_c^c(p), \label{lamdai}
\end{equation}
Hence we also find
\begin{equation}
\frac{\Gamma(1-a)^2}{\Gamma(1-2a)}=\frac{\Gamma(1+b)^2}{\Gamma(1+2b)}=\lambda_i. \label{abi}
\end{equation}
The ideal TMCT equation (\ref{K0f0}) has been solved numerically based on the Percus-Yevick model \cite{kimura16,kimu14}. Although Eq.~(\ref{lamdai}) is not solved yet, by using those numerical results, one can estimate its value. In fact, one can find $a\simeq 0.3128$ and $b\simeq 0.5869$, leading to $\lambda_i\simeq 0.7326$, where $f_c^c\simeq 0.973$.
Hence it is expected that $\lambda_i<\lambda_{mct}$ for the PY model. 

We next discuss the TMCT. Utilizing Eq.~(\ref{Kzr}) leads to
\begin{equation}
\frac{1}{zK_c[q,z]}=z\frac{S_c(q)}{q^2v_{th}^2}M_c[q,z]. \label{tmctz}
\end{equation}
Similarly to the ideal TMCT, the same equation as Eq.~(\ref{sol}) is also found from Eq.~(\ref{tmctz}), except that $\lambda_i$ is now replaced by $\lambda_{tmct}$ which is given by
\begin{equation}
\lambda_{tmct}=\frac{\lambda_i-\Delta_1}{1+\Delta_2}, \label{lamdat}
\end{equation}
where $\Delta_n$ is a positive constant and is given by
 \begin{equation}
\begin{split}
\Delta_1&= \frac{1}{2}\sum_{q,k,p}V_c^{(2)}(q,k,p)f_c^c(k)f_c^c(p)\\
&\times (1-\frac{1}{2^{1+b}})K_c^c(q)^2K_c^c(k)K_c^c(p),
\end{split}
\label{d1}
\end{equation}
\begin{equation}
\begin{split}
\Delta_2&=\frac{1}{2}\sum_{q,k,p}K_c^c(q)V_c^{(2)}(q,k,p)f_c^c(k)f_c^c(p)K_c^c(k)K_c^c(p)\\
&\times \bigg[K_c^c(q)\frac{\Gamma[2+2b]}{(b+1)\Gamma[1+b]^2}
+(K_c^c(k)+K_c^c(p))\\
&\times \Big(1-K_c^c(q)K_c^c(q)\frac{\Gamma[2+2b]}{(1+b)\Gamma[1+b]^2}\Big)\\
&-\frac{1}{2^{1+b}}K_c^c(q)K_c^c(k)K_c^c(p)\bigg].
\end{split}
\label{d2}
\end{equation}
Here it should be noted that the correction terms $\Delta_n$ appear because the memory function $S_c(q)M_c[q,z]/(q^2v_{th}^2)$ differs from $\mathcal{L}[\mathpzc{F}_c(q,f_c(t),f_c(t))][z]$ due to the non-Markovian factor $f_{\alpha}(q,t)/f_{\alpha}(q,s+t) (>1)$ in Eq.~(\ref{mz}). The asymptotic behavior obtained by the ideal MCT and the ideal TMCT is shown to deviate numerically from the simulation results and the experimental data in the actual $\beta$ stage. In fact, both theories give $\alpha_2(t)<0$ in the $\beta$ stage. On the other hand, the TMCT gives $\alpha_2(t)>0$ in the $\beta$ stage because of the non-Markovian factor \cite{toku17}. Hence the exponent parameter $\lambda_{tmct}$ is expected to satisfy the relation given by  $\lambda_{mct}>\lambda_i>\lambda_{tmct}$. In fact, the simulation results for the hard spheres with 15$\%$ polydispersity \cite{ku06} and also for the soft spheres with 15$\%$ polydispersity \cite{n08} suggest that $a\simeq 0.3177$ and $b\simeq 0.6051$, leading to $\lambda\simeq 0.7215$ \cite{toku192}, where $\lambda$ is given by $\lambda=\Gamma(1-a)^2/\Gamma(1-2a)=\Gamma(1+b)^2/\Gamma(1+2b)$. To obtain the precise asymptotic behavior of $f_c(q,t)$ in the $\beta$ stage, therefore, one must solve Eq.~(\ref{K0f}) numerically. Since such a numerical analysis is beyond the scope of the present study, it will be addressed in a future publication.

\subsection{A stretched exponential decay in an $\alpha$ stage}
At the so-called $\alpha$-relaxation stage following the $\beta$ stage, $f_c(q,t)$ is also known to obey the Kohlrausch-Williams-Watts (KWW) function, which is often referred to as the stretched exponential function, i.e., 
\begin{equation}
f_c(q,t)=f_c^c(q)\exp\left[-(t/\tau_{\alpha})^{\beta_{KWW}}\right], \label{KWW}
\end{equation}
where $\beta_{KWW}$ is the stretched exponent and $\tau_{\alpha}$ is the $\alpha$-relaxation time. 

The numerical solutions of the full TMCT equations have been shown to be well described by the same asymptotic laws as those obtained from the ideal TMCT near $\xi_c$ in the $\alpha$ stage. This is because Eq.~(\ref{K0f}) asymptotically reduces to Eq.~(\ref{K0f0}) in such a long-time regime. By utilizing the Percus-Yevick static structure factor, previous numerical analysis found $\beta_{KWW} \simeq 0.727$ for the ideal TMCT, whereas $\beta_{KWW} \simeq 0.66$ for MCT \cite{kimura16,kimu14}.

\section{Self-diffusion process}
In this section, we discuss the dynamics of a tagged particle and show how the new formulation for $\alpha_2(t)$ obtained from Eq.~(\ref{Kcumu}) [or Eq.~(\ref{K0f})] differs from the previous expression obtained from Eq.~(\ref{Kcumu0}) [or Eq.~(\ref{K0f0})]. 

By employing the cumulant expansion method \cite{kubo62}, the self-intermediate scattering function $f_s(q,t)$ can be written as
\begin{equation}
f_s(q,t)=\exp\left[-\frac{q^2}{6}M_2(t)+\frac{q^4}{2}\left(\frac{M_2(t)}{6}\right)^2\alpha_2(t)+\cdots\right], \label{cumus}
\end{equation}
with the non-Gaussian parameter defined by
\begin{equation}
\alpha_2(t)=\frac{3M_4(t)}{5M_2(t)^2}-1,\label{al2}
\end{equation}
where $M_{2n}(t)=\langle|\bm{X}_i(t)-\bm{X}_i(0)|^{2n}\rangle$. Expanding $K_s(q,t)$ in powers of $q$ as
\begin{equation}
K_s(q,t)=q^2K_2(t)-q^4K_4(t) +\cdots, \label{Kexp}
\end{equation}
we obtain
\begin{equation}
\begin{split}
M_2(t) &= 6K_2(t), \quad M_4(t)=120K_4(t)+60K_2(t)^2, \\
\alpha_2(t) &= 2\frac{K_4(t)}{K_2(t)^2}.
\end{split}
\label{Mn_al2k}
\end{equation}

To find the expansion of the self nonlinear memory function $\Delta\varphi_s(q,t)$ in powers of $q$, we first expand the self-intermediate scattering function $f_s(|\bm{q}-\bm{k}|,t)$ as
\begin{equation}
\begin{split}
&f_s(|\bm{q}-\bm{k}|,t) = f_s(k,t)-q(\hat{\bm{q}}\cdot\hat{\bm{k}})\frac{\partial}{\partial k}f_s(k,t) \\
&+ \frac{q^2}{2}\left[(\hat{\bm{q}}\cdot\hat{\bm{k}})^2\frac{\partial^2}{\partial k^2}f_s(k,t) + \frac{1-(\hat{\bm{q}}\cdot\hat{\bm{k}})^2}{k}\frac{\partial}{\partial k}f_s(k,t)\right]\\
&+ O(q^4).
\end{split}
\label{fsexpan}
\end{equation}
From Eq.~(\ref{memory}), we then obtain
\begin{equation}
\Delta\varphi_s(q,t)=\Delta\varphi_s^{(0)}(t)+q^2\Delta\varphi_s^{(2)}(t)+\cdots, \label{memoexp}
\end{equation}
with
\begin{equation}
\Delta\varphi_s^{(0)}(t) = \frac{\rho v_{th}^2}{6\pi^2}\int_0^{q_c}dk k^4c(k)^2S(k)f_c(k,t)f_s(k,t), \label{memo0}
\end{equation}
\begin{equation}
\begin{split}
\Delta\varphi_s^{(2)}(t) &= \frac{\rho v_{th}^2}{20\pi^2}\int_0^{q_c}dk k^4c(k)^2S(k)f_c(k,t)\\
&\times \left(\frac{2}{3k}\frac{\partial}{\partial k}+\frac{\partial^2}{\partial k^2}\right)f_s(k,t). 
\end{split}
\label{memo02}
\end{equation}
\begin{table*}[!t]
\caption{Main differences between TMCT and MCT.}
\begin{singlespace}
\centering
\resizebox{\textwidth}{!}{
\begin{tabular}{cccc}
\hline
Theory & TMCT &MCT\\
\hline
Stage [K] & &\\
a linear equation for $\psi_{\alpha}(q,t)$ & time-convolution & time-convolution \\
with memory function & $\Delta\varphi_{\alpha}(q,t)$ & $\Delta\varphi_{\alpha}(q,t)$\\
a linear equation for $f_{\alpha}(q,t)$ & "time-convolutionless" & "time-convolution"\\
with memory function & $\psi_{\alpha}(q,t)$ & $\psi_{\alpha}(q,t)$\\
\hline
Stage [H] & TMCT equation for $K_{\alpha}(q,t)$&----- \\
non-Gaussian parameter & $\alpha_2(t=0)=\alpha_2(t=\infty)=0$ & -----\\
$\alpha_2(t)$ & $\alpha_2(t)\geq 0$  $(0\leq t\leq \infty)$&\\
\hline
Stage [H$_2$]&    ideal TMCT equation for $K_{\alpha}(q,t)$ & ideal MCT equation for $f_{\alpha}(q,t)$\\
$\alpha_2(t)$ & $\alpha_2(t=0)=0$ & $\alpha_2(t=0)=-2/3$\\
                      & $\alpha_2(t=\infty)=0$ & $\alpha_2(t=\infty)=0$\\
$\beta$-relaxation stage&             $\alpha_2(t)<0$       &   $\alpha_2(t)<0$\\
\hline
non-ergodicity parameter& &\\
{\Large$f_{\alpha}(q)$} & {\LARGE $e^{- \frac{1}{\mathpzc{F}_{\alpha}(q)}}$} & {\LARGE$\frac{\mathpzc{F}_{\alpha}(q)}{1+\mathpzc{F}_{\alpha}(q)}$}\\
\hline
critical volume fraction $\phi_c$&&\\
for Percus-Yevick model$^{(a)}$  & 0.5856 ($q_c\sigma$=20), 0.5817 ($q_c\sigma$=40) &$\quad$ 0.5214 ($q_c\sigma$=20), 0.5159 ($q_c\sigma$=40)\\
\hline
(a) Refs. \cite{kimura16,kimu14}.                     
\end{tabular}}
\label{table-1}
\end{singlespace}
\end{table*}

Utilizing Eqs.~(\ref{K0f}), (\ref{Kexp}), and (\ref{memoexp}) thus leads to
\begin{equation}
K_2(t) = v_{th}^2C_s(t)-\int_0^t\Gamma_0(s)K_2(t-s)ds, \label{K2sf}
\end{equation}
\begin{equation}
\begin{split}
K_4(t) &= -\int_0^t\Gamma_0(s)K_4(t-s)ds + \int_0^t\Gamma_2(s)K_2(t-s)ds \\
&+ w\int_0^t ds\int_0^{t-s}d \tau  e^{-\gamma_{\alpha}(t-s-\tau)}\Delta\varphi_s^{(0)}(\tau)\\
&\times \left[K_2(s+\tau)-K_2(s)\right]K_2(s),
\end{split}
\label{K4sf}
\end{equation}
where
\begin{equation}
\Gamma_n(t)=\int_0^te^{-\gamma_{\alpha} (t-s)}\Delta\varphi_s^{(n)}(s)ds. \label{Ga0}
\end{equation}
Here, $w$ is a parameter introduced to distinguish the theories, where $w=1$ corresponds to the full TMCT and $w=0$ denotes the ideal TMCT. 

Next, we derive the formal equations for $M_{2n}(t)$. Combining Eqs.~(\ref{Mn_al2k}), (\ref{K2sf}), and (\ref{K4sf}) yields
\begin{equation}
M_2(t) = 6v_{th}^2C_s(t)-\int_0^t\Gamma_0(t-s)M_2(s)ds, \label{M2sf}
\end{equation}
\begin{equation}
\begin{split}
M_4(t) &= \frac{5}{3}M_2(t)^2-\int_0^t\Gamma_0(t-s)\left[M_4(s)-\frac{5}{3}M_2(s)^2\right]ds \\
&+ 20\int_0^t\Gamma_2(t-s)M_2(s)ds \\
&+ w\int_0^t ds\int_0^{t-s}d \tau  e^{-\gamma_{\alpha}(t-s-\tau)}\Delta\varphi_s^{(0)}(\tau)\\
&\times \left[M_2(s+\tau)-M_2(s)\right]M_2(s).
\end{split}
\label{M4sf}
\end{equation}
It is worth noting that the last terms in Eqs.~(\ref{K4sf}) and (\ref{M4sf}) with $w=1$ play a crucial role in ensuring a physically reasonable behavior of $\alpha_2(t)$ for intermediate times, leading to $\alpha_2(t)\geq 0$ for $0\leq t\leq \infty$. Without these terms (i.e., setting $w=0$), $K_4(t)$ and $M_4(t)$ become negative at intermediate times, which leads to an unphysical $\alpha_2(t)<0$ in the $\beta$ stage. This occurs because the second term of Eq.~(\ref{K4sf}) and the third term of Eq.~(\ref{M4sf}), which contain the nonlinear memory function $\Delta \varphi_s^{(2)}(t)$, become negative in this time region \cite{toku17}. 

Finally, we discuss the asymptotic behavior in both theories. As mentioned above, the asymptotic behavior of $M_{2n}(t)$ in the full TMCT is identical to that in the ideal TMCT. In the short-time limit $t\rightarrow 0$, one obtains
\begin{equation}
M_2(t)\simeq 3v_{th}^2t^2, \quad M_4(t)\simeq\frac{5}{3}M_2(t)^2. \label{m4st}
\end{equation}
In the long-time limit $t\rightarrow \infty$, one obtains for $\xi<\xi_c$:
\begin{equation}
M_2(t)\simeq 6D_s^Lt, \quad M_4(t)\simeq\frac{5}{3}M_2(t)^2, \label{m4lt}
\end{equation}
with the long-time self-diffusion coefficient given by
\begin{equation}
D_s^L=\frac{v_{th}^2}{\gamma_{\alpha}+\int_0^{\infty}\Delta\varphi_s^{(0)}(s)ds}. \label{stsdc}
\end{equation}
From Eqs.~(\ref{al2}), (\ref{m4st}), and (\ref{m4lt}), it is straightforward to show that $\alpha_2(0)=\alpha_2(\infty)=0$. Consequently, the difference between the two theories regarding $M_{2n}(t)$ manifests itself exclusively at intermediate times.

\section{Main differences between TMCT and MCT}
In the present section, we discuss the main differences between TMCT and MCT, which are summarized in Table~\ref{table-1}. In stage [K], the governing equation for $\psi_{\alpha}(q,t)$ takes a time-convolution form for both TMCT and MCT, although the approximation (A2) is additionally used in MCT. Here, we note that both equations are expressed in terms of the same memory function $\Delta\varphi_{\alpha}(q,t)$. 

The most fundamental difference is found in the linear equation for $f_{\alpha}(q,t)$. This equation is time-convolutionless for TMCT [see Eq.~(\ref{eqf})], whereas it takes a time-convolution form for MCT [see Eq.~(\ref{mct0})]. As discussed in the Introduction, however, the TC formulation is not applicable to the number densities because they constitute the relevant variables in stage [H]. Consequently, this difference leads to an unphysical behavior in the physical quantities predicted within stage [H] by MCT. 

Firstly, the non-Gaussian parameter $\alpha_2(t)$ yields an unphysical value of $\alpha_2(t=0)=-2/3$ in the ideal MCT, whereas TMCT ensures a physically reasonable behavior for all times \cite{toku17}. Secondly, the non-ergodicity parameter $f_{\alpha}(q)$ obeys different equations in the two theories, resulting in distinct critical points $\xi_c$. In fact, the critical volume fraction $\phi_c$ obtained from TMCT is much higher than that predicted by MCT, irrespective of the choice of $q_c$. 

Thus, we confirm that only the TCL formulation is applicable to describing the dynamics of the number density. This stems mainly from the fact that the space-time scale of the number density is much longer than that of the current density; hence, the dynamics in stage [H] is exclusively governed by the number density. Conversely, in stage [K], only the TC formulation is applicable to the current density, which governs the early-stage dynamics of the system. 

Finally, we should mention that the unphysical behavior given by $\alpha_2(t)<0$ in the $\beta$ stage is obtained in both the ideal MCT and ideal TMCT equations, because the approximation (A2) is employed in deriving both idealized versions.

\section{Summary}
In the present paper, we have clarified how two types of projection operator methods play distinct, crucial roles in each relaxation stage. Depending on the space-time scales (or stages), the relevant physical variables change. The most important conclusion is that only the TCL formulation is applicable to the relevant variables in stage [H] (the diffusion stage), whereas the TC formulation cannot be applied. 

In stage [K], the dynamics are described by both the current and number densities, while in stage [H], the number densities exclusively become the relevant variables. Since the space-time scale over which the number densities dominate the system dynamics is much longer than that of the current densities, the TC formulation is appropriate only for the current densities, which govern the short-time dynamics within stage [K]. Conversely, the number densities play a primary role in stage [H]. Hence, one must conclude that the TCL formulation is mandatory for describing $f_{\alpha}(q,t)$ in stage [H], whereas employing the TC formulation for $f_{\alpha}(q,t)$ causes serious unphysical behaviors in various physical quantities. Numerical calculations performed in both TMCT and MCT have already strongly supported this conclusion. 

Therefore, from this point of view, it is also highly understandable why the TCL-type master equation derived from the Schr\"{o}dinger equation first proposed by Tokuyama and Mori  \cite{toku76} plays a significant role even in open quantum systems \cite{akw20,nbw21}, where particle diffusion fundamentally dominates the system dynamics.

Finally, we remark that Fig.~\ref{evo} presented in this paper provides a consolidated and updated framework for stages [K] and [H], comprehensively replacing the individual schematic representations previously plotted in Refs.~\cite{kimura16,toku191,toku18,toku141,toku19}.

\appendix
\section{Derivation of Eq. (\ref{K0f})}
In order to show how to derive Eq. (\ref{K0f}) from Eq. (\ref{Kcumu}), we start from the following equations:
\begin{equation}
\frac{\partial^2}{\partial t^2}K_{\alpha}(q,t)=q^2\psi_{\alpha}(q,t), \label{keq2}
\end{equation}
\begin{equation}
\begin{split}
\frac{\partial}{\partial \tau}\psi_{\alpha}(q,t)&=-\gamma_{\alpha} \psi_{\alpha}(q,t)\\
&-\int_0^{t}\Delta\varphi_{\alpha}(q,t-s)\frac{f_{\alpha}(q,s)}{f_{\alpha}(q,t)}\psi_{\alpha}(q,s)ds.
\end{split}
\label{yeq1}
\end{equation}
The formal solution of Eq. (\ref{yeq1}) is given by
\begin{equation}
\psi_{\alpha}(q,t)=e^{-\gamma_{\alpha}t}\psi_{\alpha}(q,0)- B_{\alpha}(q,t),\label{yfs}
\end{equation}
\begin{equation}
\begin{split}
B_{\alpha}(q,t)&=\int_0^{t}ds\frac{e^{-\gamma_{\alpha}(t-s)}}{f_{\alpha}(q,s)}\int_0^{t}dx\theta(s-x)\Delta\varphi_{\alpha}(q,s-x)\\
&\times f_{\alpha}(q,x)\psi_{\alpha}(q,x)\\
&=\int_0^{t}dsf_{\alpha}(q,s)\psi_{\alpha}(q,s)A_{\alpha}(q,t,s), 
\end{split}
\label{b2}
\end{equation}
where
\begin{equation}
A_{\alpha}(q,t,s)=\int_0^{t-s}d\tau e^{-\gamma_{\alpha}\tau}\frac{\Delta\varphi_{\alpha}(q,t-s-\tau)}{f_{\alpha}(q,t-\tau)}. \label{a1}
\end{equation}
Use of Eqs. (\ref{keq2}) and (\ref{yfs}) then leads to
\begin{equation}
\begin{split}
\frac{\partial}{\partial \tau}K_{\alpha}(q,t)&=q^2(1-e^{-\gamma_{\alpha}t})\psi_{\alpha}(q,0)/\gamma_{\alpha}\\
&- q^2\int_0^{t}B_{\alpha}(q,s)ds,
\end{split}
\label{ks1}
\end{equation}
\begin{equation}
K_{\alpha}(q,t)=\frac{q^2v_{th}^2}{S_{\alpha}(q)}C_{\alpha}(t)-H_{\alpha}(q,t),\label{ks}
\end{equation}
where
\begin{equation}
H_{\alpha}(q,t)=q^2\int_0^{t}(t-s)B_{\alpha}(q,s)ds. \label{h1}
\end{equation}
By using Eq. (\ref{keq2}), one can rewrite $H_{\alpha}(q,t)$ as
\begin{equation}
\begin{split}
&H_{\alpha}(q,t)\\
&=\int_0^{t}ds(t-s)\int_0^sdxf_{\alpha}(q,x)\frac{d^2K_{\alpha}(q,x)}{dx^2}A_{\alpha}(q,s,x)\\
&=\int_0^{t}dxf_{\alpha}(q,x)\frac{d^2K_{\alpha}(q,x)}{dx^2}\int_0^{t-x}ds sA_{\alpha}(q,t-s,x)\\
&=[f_{\alpha}(q,x)\frac{dK_{\alpha}(q,x)}{dx}\int_0^{t-x}ds sA_{\alpha}(q,t-s,x)]_0^{t}\\
&-\int_0^{t}dx\frac{f_{\alpha}(q,x)}{dx}\frac{dK_{\alpha}(q,x)}{dx}\int_0^{t-x}sA_{\alpha}(q,t-s,x)ds\\
&-\int_0^{t}dxf_{\alpha}(q,x)\frac{dK_{\alpha}(q,x)}{dx}\frac{d}{dx}\int_0^{t-x}sA_{\alpha}(q,t-s,x)ds\\
\end{split}
\label{h2}
\end{equation}
\newpage %
Neglecting the terms containing $D(q,x)$, one can obtain
\begin{equation}
\begin{split}
&H_{\alpha}(q,t)\simeq\int_0^{t}dxf_{\alpha}(q,x)\frac{dK_{\alpha}(q,x)}{dx}\int_0^{t-x}dsA_{\alpha}(q,s+x,x)\\
&=[f_{\alpha}(q,x)K_{\alpha}(q,x)\int_0^{t-x}dsA_{\alpha}(q,s+x,x)]_0^{t}\\
&-\int_0^{t}dx\frac{df_{\alpha}(q,x)}{dx}K_{\alpha}(q,x)\int_0^{t-x}dsA_{\alpha}(q,s+x,x)\\
&+\int_0^{t}f_{\alpha}(q,s)K_{\alpha}(q,s)A_{\alpha}(q,t,s)ds\\
&\simeq\int_0^{t}f_{\alpha}(q,s)K_{\alpha}(q,s)A_{\alpha}(q,t,s)ds.
\end{split}
\label{h3}
\end{equation}
Use of Eqs. (\ref{ks}) and (\ref{h2}) then leads to Eq. (\ref{K0f}).


\begin{references}
\bibitem{mct84B}U. Bengtzelius, W. G\"{o}tze, A. Sj\"{o}lander, J. Phys. C {\bf17} (1984) 5915.
\bibitem{mct84L}E. Leutheusser, Phys. Rev. A {\bf29} (1984) 2765.
\bibitem{toku14}M. Tokuyama, Physica A {\bf395} (2014) 31.
\bibitem{toku17}M. Tokuyama, Physica A {\bf484} (2017) 453.
\bibitem{mori65}H. Mori, Prog. Theor. Phys. {\bf33} (1965) 423.
\bibitem{toku75}M. Tokuyama and H. Mori, Prog. Theor. Phys. {\bf54} (1975) 918.
\bibitem{toku76}M. Tokuyama and H. Mori, Prog. Theor. Phys. {\bf55} (1976) 411.
\bibitem{pusey81}G. L. Paul, P. N. Pusey, J. Phys. A {\bf14} (1989) 3301.
\bibitem{hu11}R. Huang, I. Chavez, K. M. Taute, B. Lukić, S. Jeney, M. G. Raizen, Ernst-Ludwig Florin, Nature Physics {\bf7} (2011) 422.
\bibitem{toku78} M. Tokuyama, I. Oppenheim, Physica A {\bf94} (1978) 501.
\bibitem{toku761}M. Tokuyama and H. Mori, Prog. Theor. Phys. {\bf56} (1976) 1073.
\bibitem{cohen}E. G. D. Cohen and T. H. Berlin, Physica {\bf26} (1960) 717.

\bibitem{mori73}Prog. Theor. Phys. {\bf49} (1973) 1516.
\bibitem{toku81}M. Tokuyama, Physica A {\bf109} (1981) 128.
\bibitem{toku15}M. Tokuyama, Physica A {\bf 430} (2015) 156.
\bibitem{mct91}W. G\"{o}tze, in: J.P. Hansen, D. Levesque, J. Zinn-Justen (Eds.), Liquids, Freezing and Glass Transition, North-Holland, Amsterdam, 1991.
\bibitem{fuch} M. Fuchs, W. G\"{o}tze, I. Hofacker, and A. Latz, J. Phys. Condens. Matter {\bf3} (1991) 5047.
\bibitem{fuchs}M. Fuchs, Transport Theory and Statistical Physics {\bf24} (1995) 855.
\bibitem{fran}T. Franosch, M. Fuchs, W. G\"{o}tze, M. R. Mayr, and A. P. Singh, Phys. Rev. E {\bf55} (1997) 7153.
\bibitem{chong} S.-H. Chong, W. G\"{o}tze, and M. R. Mayr, Phys. Rev. E {\bf64} (2001) 011503.
\bibitem{voigt03}Th. Voigtmann, Phys. Rev. E {\bf68} (2003) 051401.
\bibitem{got03}W. G\"{o}tze and Th. Voigtmann, Phys. Rev. E {\bf67} (2003) 021502.

\bibitem{sza}G. Szamel, Phys. Rev. Lett. {\bf90} (2003) 228301.
\bibitem{foffi}G. Foffi, W. G\"{o}tze, F. Sciortino, P. Tartaglia, and Th. Voigtmann, Phys. Rev. E {\bf69} (2004) 011505.
\bibitem{voigt04}Th. Voigtmann, A. M. Puertas, and M. Fuchs, Phys. Rev. E {\bf70} (2004) 061506.
\bibitem{flenn}E. Flenner and G. Szamel, Phys. Rev. E {\bf72} (2005) 031508.
\bibitem{voigt06}Th. Voigtmann and J. Horbach, Europhys. Lett. {\bf74} (2006) 459.
\bibitem{toku08}M. Tokuyama and Y. Kimura, Physica A {\bf387} (2008) 4749.
\bibitem{got09} W. G\"{o}tze, {\it Complex Dynamics of Glass Forming Liquids: A Mode Coupling Theory} (Oxford Science, Oxford, 2009).
\bibitem{voigt10}F. Weysser, A. M. Puertas, M. Fuchs, and Th. Voigtmann, Phys. Rev. E {\bf82} (2010) 011504.
\bibitem{narumi11}T. Narumi and M. Tokuyama, Phys. Rev. E {\bf84} (2011) 022501.
\bibitem{voigt11}M. Domschke, M. Marsilius, T. Blochowicz, and Th. Voigtmann, Phys. Rev. E {\bf84} (2011) 031506.

\bibitem{debe}P. G. Debenedetti and F. H. Stillinger, Nature {\bf410} (2001) 259.
\bibitem{toku171}M. Tokuyama, Physica A. {\bf465} (2017) 229.
\bibitem{kimura16}Y. Kimura and M. Tokuyama, IL Nuovo Cimento C {\bf39} (2016) 300.
\bibitem{kimu14}Y. Kimura and M. Tokuyama, to be submitted to Physica A.
\bibitem{PY}J. K. Percus and G. J. Yevick, Phys. Rev. {\bf110} (1958) 1.
\bibitem{toku03}M. Tokuyama, H. Yamazaki, Y. Terada, Phys. Rev. E {\bf67} (2003) 062403.
\bibitem{toku031} M. Tokuyama, H. Yamazaki, Y. Terada, Physica A {\bf328} (2003) 367.
\bibitem{toku191}M. Tokuyama, R. Takekawa, and J. Kawamura, Physica A {\bf529} (2019) 121541.
\bibitem{got15}W. G\"{o}tze and R. Schilling, Phys. Rev. E {\bf91} (2015) 042117.
\bibitem{kubo62} R. Kubo, J. Phys. Soc. Japan {\bf17} (1962) 1100.
\bibitem{ng}A. Rahman, K.S. Singwi, A. Sj\"{o}lander, Phys. Rev. {\bf126} (1962) 986.

\bibitem{toku18} M. Tokuyama, T. Narumi, and J. Kawamura, Physica A {\bf512} (2018) 552.
\bibitem{toku192}M. Tokuyama, Physica A {\bf526} (2019) 121074.
\bibitem{toku05} M. Tokuyama, Y. Terada, J. Chem. Phys. B {\bf109} (2005) 21357.
\bibitem{toku071} M. Tokuyama, T. Narumi, E. Kohira, Physica A {\bf385} (2007) 439.
\bibitem{toku141}M. Tokuyama, arXiv:1409.4839 [cond-mat.stat-mech].
\bibitem{toku19}M. Tokuyama and T. Narumi, Physica A {\bf514} (2019) 533.
\bibitem{daw}K.Dawson, G. Foffi, M. Fuchs, W. G\"{o}tze, F. Sciortino, M. Sperl, P. Tartaglia, T. Voigtmann, and E. Zaccarelli, Phys. Rev. E {\bf63} (2000) 011401.
\bibitem{naru17}T. Narumi and M. Tokuyama, Phys. Rev. E {\bf95} (2017) 032601.
\bibitem{ku06} S.K. Kumar, G. Szamel, J.F. Douglas, J. Chem. Phys. {\bf124} (2006) 214501.
\bibitem{n08} S. Nakanishi, T. Narumi, Y. Terada, M. Tokuyama, Rep. Inst. Fluid Sci. {\bf} 19 (2008) 1.

\bibitem{akw20}D. Ahn, J.H. Oh, K. Kimm, and S. W. Hwang, Phys. Rev. A {\bf61} (2000) 052310.
\bibitem{nbw21}K. Nestmann, V. Bruch, and M. R. Wegewijs, Phys. Rev. X {\bf11} (2021) 021041.
\end{references}
\end{document}